\documentclass[twoside,fleqn]{article}
\usepackage{espcrc2,amssymb,hyperref}
\usepackage[dvips]{color}
\usepackage[dvips]{graphicx}

\newcommand{\bvec}[1]{\ensuremath{\mbox{\boldmath $\mathrm{#1}$}}}

\begin{document}

\title{
  Pion form factor using domain wall valence
  and \texttt{asqtad} sea quarks
}

\author{
  G.\ T.\ Fleming\address{
    Thomas Jefferson National Accelerator Facility,
    Newport News, VA 23606, USA.
  }\address{
    Sloane Physics Laboratory,
    Yale University,
    New Haven, CT, 06520, USA
  },
  F.\ D.\ R.\ Bonnet$^\mathrm{a}$\address{
    Department of Physics,
    University of Regina,
    Regina, SK, S4S 0A2, Canada.
  },
  R.\ G.\ Edwards$^\mathrm{a}$,
  R.\ Lewis$^\mathrm{c}$ and
  D.\ G.\ Richards$^\mathrm{a}$
}

\begin{abstract}
We compute the pion electromagnetic form factor in a hybrid calculation
with domain wall valence quarks and improved staggered (\texttt{asqtad})
sea quarks.  This method can easily be extended
to $\rho\to\gamma\pi$ transition form factors.
\end{abstract}

\maketitle 

The pion electromagnetic form factor is considered a good observable
for studying the onset with increasing energy of the perturbative QCD regime
for exclusive processes.  As the pion is the lightest and simplest hadron,
a perturbative description is believed to be valid at lower energy scales
than predictions for heavier and more complicated hadrons
like the nucleon \cite{Isgur:1984jm}.

\begin{figure}[ht]
\includegraphics[width=0.46\textwidth]{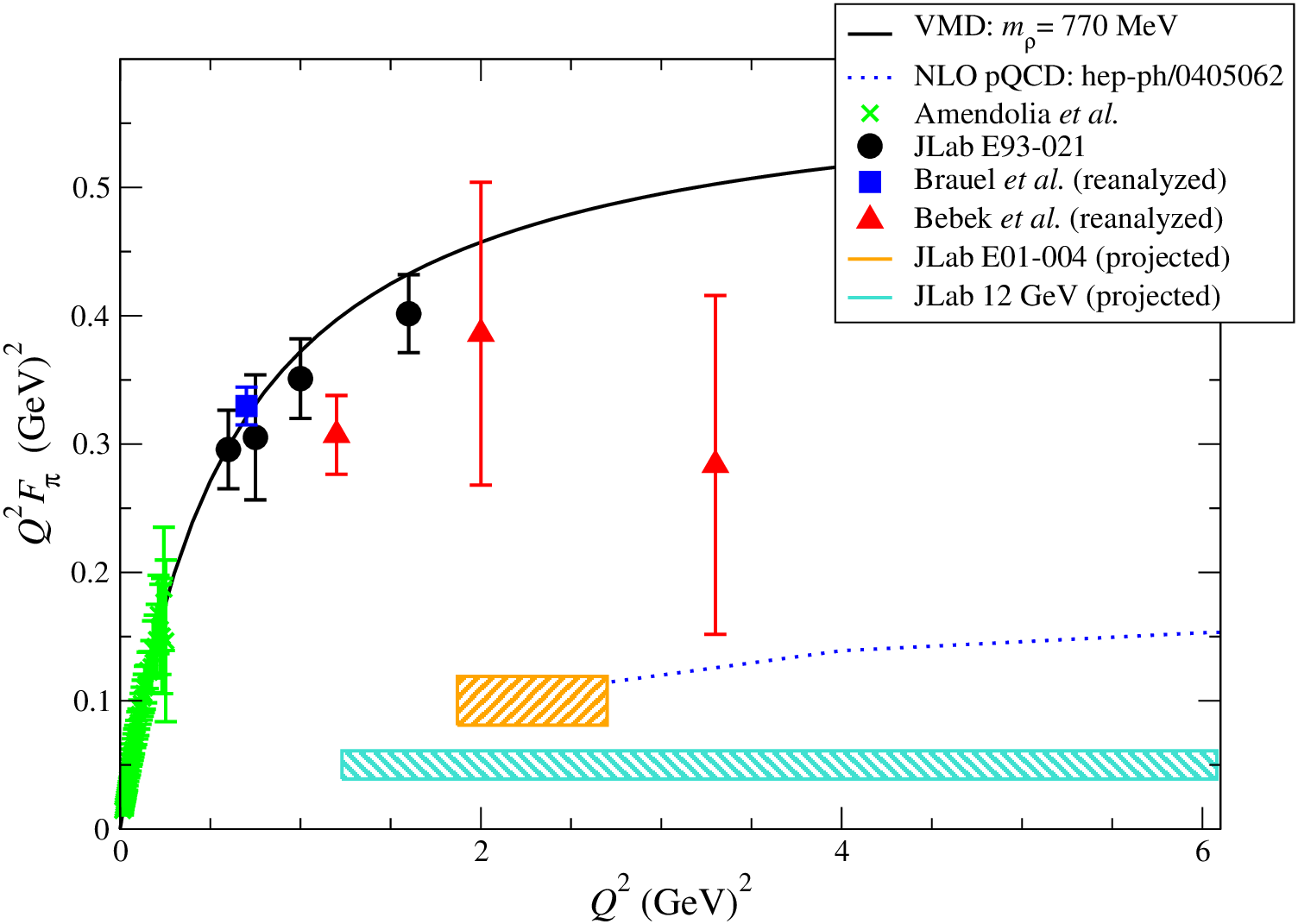}
\vspace{-4ex}
\caption{\label{fig:Fpi_expt}Summary of experimental data
for the pion electromagnetic form factor.  Shaded regions are
expected sensitivities of future experiments.}
\vspace{-4ex}
\end{figure}

A pseudoscalar particle has only a single electromagnetic form factor,
$F(Q^2)$, where $Q^2$ is the four-momentum transfer,
and at $Q^2 = 0$, this form factor is the electric charge
of the particle, $F(Q^2 = 0) = 1$; the magnetic form factor vanishes.
The experimentally observed behavior of the form factor
at small momentum transfer is well described by the vector meson dominance
(VMD) hypothesis \cite{Holladay:1955,Frazer:1959gy,Frazer:1959}
\begin{equation}
\label{eq:vmd_form}
F_\pi(Q^2) \approx \frac{1}{1+Q^2 \left/ m_\mathrm{VMD}^2 \right.}
\quad \mathrm{for} \quad Q^2 \ll m_\mathrm{VMD}^2
\end{equation}
The current experimental situation is presented in Fig.~\ref{fig:Fpi_expt}
\cite{Amendolia:1986wj,Bebek:1978pe,Brauel:1979zk,Volmer:2000ek,Blok:2002ew}.

What is surprising is that VMD with only the lightest resonance
($m_\rho = m_\mathrm{VMD}$) appears to describe all the existing data,
even up to scales of $Q^2 \gtrsim 1 \mathrm{GeV}^2$.  In contrast,
at very high momentum transfer, we expect the data
to approach the perturbative behavior
\cite{Brodsky:1973kr,Brodsky:1975vy,Farrar:1979aw}
\begin{equation}
F_\pi(Q^2) = \frac{8\pi\alpha_s(Q^2)f_\pi^2}{Q^2} \quad \mathrm{as} \quad
Q^2 \to \infty
\end{equation}
Higher order perturbative calculations of the hard contribution
to the form factor \cite{Stefanis:1998dg,Stefanis:2000vd,Bakulev:2004cu}
do not vary significantly from this value, as seen in Fig.~\ref{fig:Fpi_expt}.
At the largest energy scale where reliable experimental measurements have
so far been obtained, around $Q^2 \simeq 2~{\rm GeV}^2$,
the data are 100\% larger than this pQCD asymptotic prediction.

Early lattice calculations validated the vector meson dominance hypothesis
at low $Q^2$ \cite{Martinelli:1988bh,Draper:1989bp}. Recent lattice results
\cite{vanderHeide:2003ip,Nemoto:2003ng,vanderHeide:2003kh,Abdel-Rehim:2004sp},
including some of our own preliminary results
\cite{Bonnet:2003pf,Bonnet:2003aa}, have somewhat extended the range
of momentum transfer, up to 2 $\mathrm{GeV}^2$, and the results remain
consistent with VMD and the experimental data.

\begin{table}
  \caption{\label{eq:DWF_details}Simulation details for domain wall fermion
    calculations on 20$^3\times$64 dynamical MILC \texttt{asqtad} lattices
    at $a^{-1} \approx 1.6\ \mathrm{GeV}$.}
  \vspace{2ex}
  \begin{tabular}{ll|l|ll}
    $am_{ud}$ & $am_s$ & $am_\mathrm{val}$ & $m_\rho \mathrm{(MeV)}$ & $m_\pi \mathrm{(MeV)}$ \\
    \hline
    0.01 & 0.05 & 0.01  &  956(22) & 318(3) \\
    0.05 & 0.05 & 0.05  &  955(19) & 602(5) \\
    0.05 & 0.05 & 0.081 & 1060(14) & 758(5)
  \end{tabular}
\end{table}

\begin{figure}
  \begin{center}
    \includegraphics[width=0.45\textwidth]{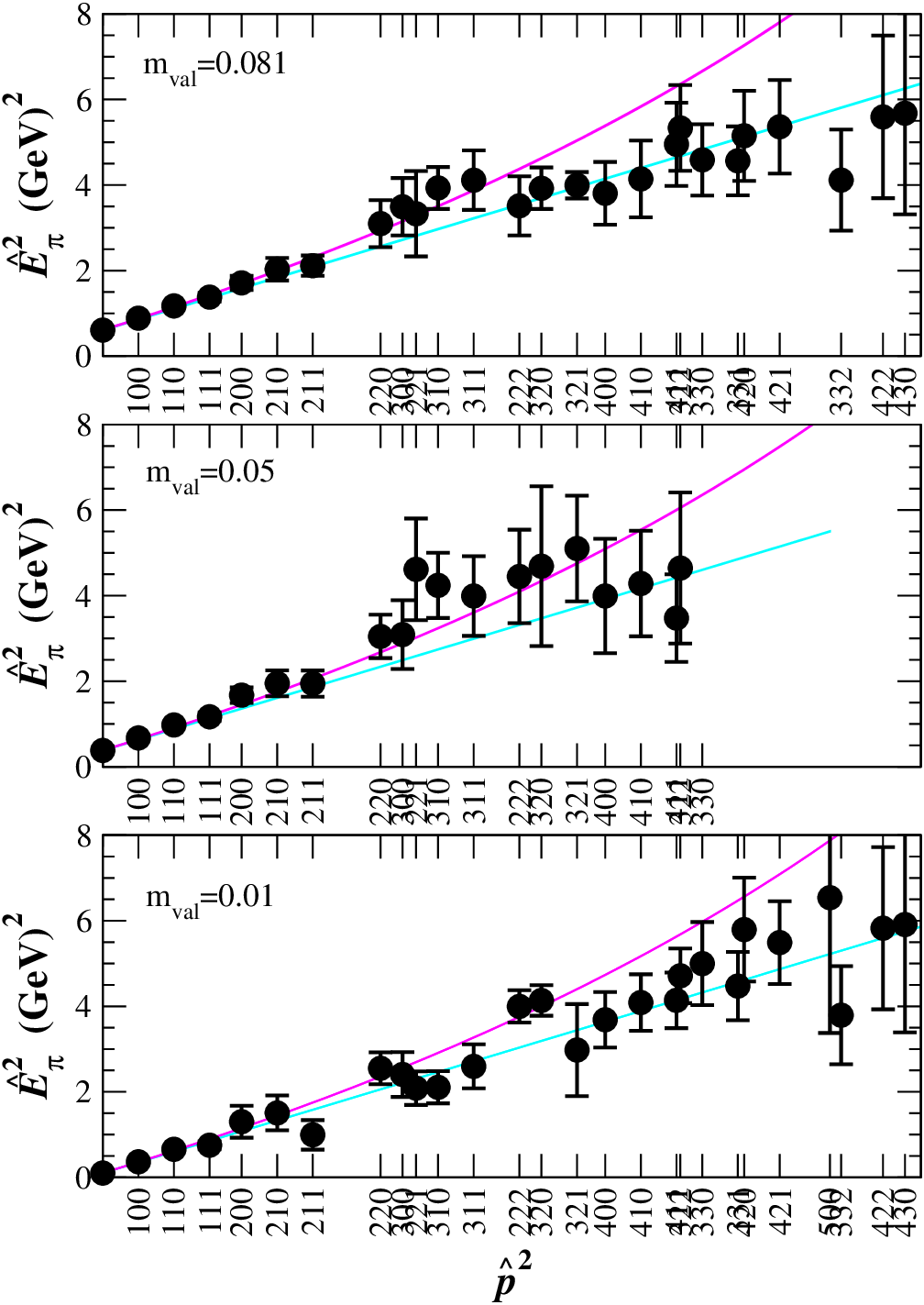}
  \end{center}
  \caption{\label{fig:dwf_dispersion}Pion (left) and rho meson (right)
    dispersion relation \textit{vs.}\ continuum (upper) and lattice (lower)
    expectations curves.}
\end{figure}

For our exploration of the pion form factor with unquenched gauge
configurations, we performed a hybrid calculation using MILC
$N_f=2+1$ and $N_f=3$ configurations in 20$^3\times$64 volumes, generated
with staggered \texttt{asqtad} sea quarks\cite{Bernard:2001av},
and domain wall valence quarks with domain wall height $m_0=1.7$
and extent $L_s = 16$ of the extra dimension \cite{Negele:2004iu}.
The MILC configurations were HYP blocked \cite{Hasenfratz:2001hp}
before valence propagators were computed, otherwise the residual
chiral symmetry breaking would have been unacceptably large.
Dirichlet boundary conditions were imposed 32 timeslices apart.
Thus, opposite halves of the lattices were used
on alternate configurations in the Monte Carlo sequence in order to reduce
autocorrelations.  A detailed study of the physical properties of light
hadrons composed of staggered quarks computed on these lattices has
recently been completed\cite{Aubin:2004wf}.  Details of the observables
we computed are available in our earlier references
\cite{Bonnet:2003pf,Bonnet:2003aa}.

To compute the pion form factor at large momentum transfer,
we must first understand the pion dispersion relation
when the momenta are relatively large.  In the continuum limit,
we expect proper relativistic behaviour
\begin{equation}
\label{eq:continuum_dispersion}
E^2_\pi(\bvec{p}) = \bvec{p}^2 + E^2_\pi(0).
\end{equation}
From the study of free lattice bosons, we can define
lattice equivalents of the continuum momentum and energy
\begin{equation}
\widehat{E} = 2 \sinh \left( E / 2 \right) \qquad
\widehat{p}_x = 2 \sin  \left( p_x / 2 \right).
\end{equation}
Another possibility is that the pion dispersion relation
will follow that of a free lattice boson
\begin{equation}
\label{eq:free_lattice_boson_dispersion}
\widehat{E}^2_\pi(\bvec{\widehat{p}})
  = \bvec{\widehat{p}}^2 + \widehat{E}^2_\pi(0)
\end{equation}
Eqs.~(\ref{eq:continuum_dispersion})
and (\ref{eq:free_lattice_boson_dispersion})
differ significantly only when the lattice momenta are large.

In Fig.~\ref{fig:dwf_dispersion}, we have plotted against each
dispersion relation and we see that the data clearly favor
the lattice dispersion relation at large momenta.  This gives us confidence
that we can use Eq.~(\ref{eq:free_lattice_boson_dispersion}) to determine
$E_\pi(\bvec{p})$ at high momenta from fits to low momentum correlators
which have much less statistical noise.

\begin{figure}
\includegraphics[width=0.46\textwidth]{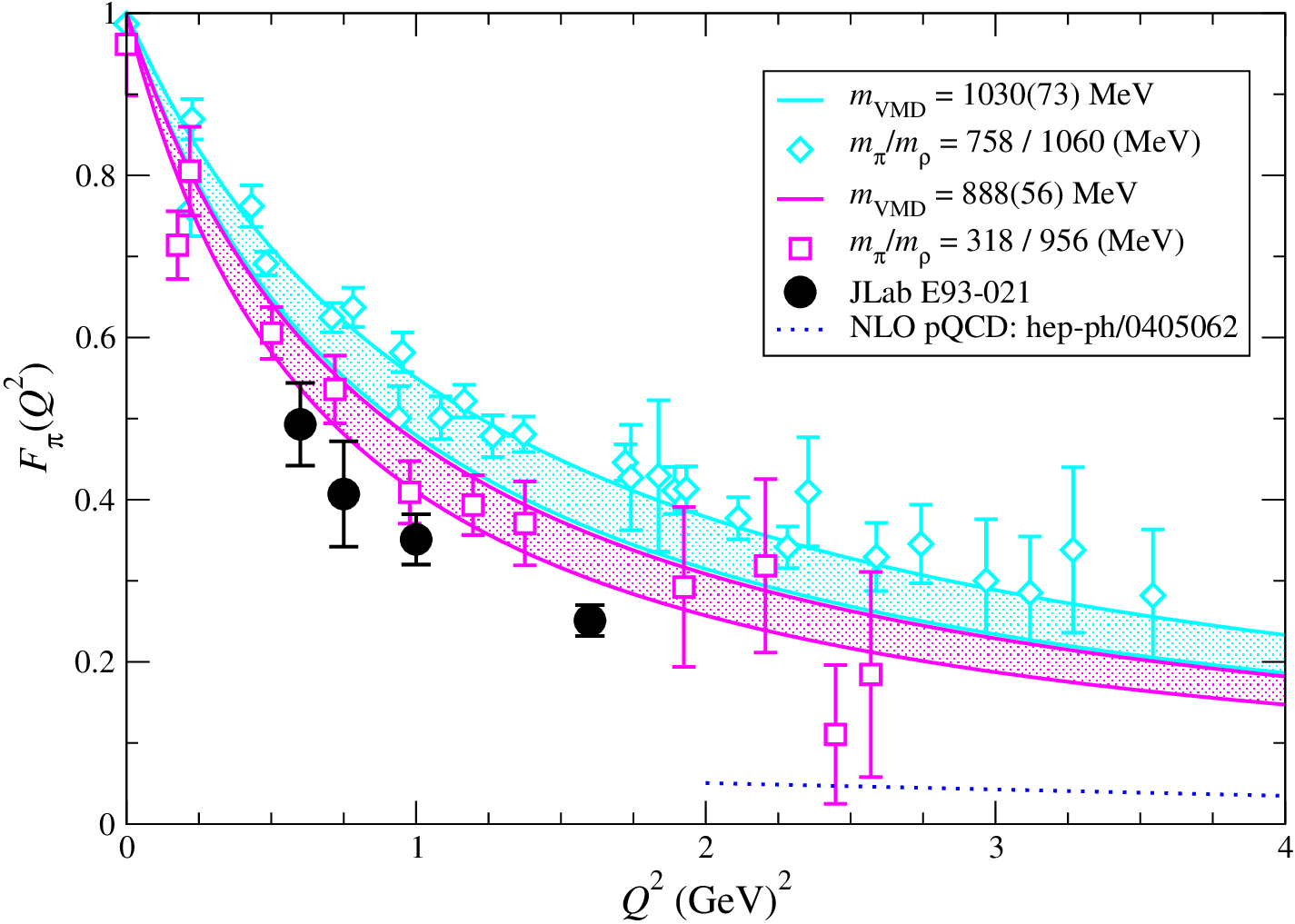}
\vspace{-5ex}
\caption{\label{fig:F_px1_py0_pz0} Pion electromagnetic form factor
  for fixed sink momentum $\bvec{p}_f = (1,0,0)$ computed by the ratio method
  \cite{Bonnet:2003pf,Bonnet:2003aa} and imposing the lattice dispersion
  relation, Eq.~(\ref{eq:free_lattice_boson_dispersion}). Shaded regions
  are jackknife error bands for VMD fit.
}
\vspace{-5ex}
\end{figure}

Using the lattice dispersion relation and the ratio method described
in our previous work, we have computed in Fig.~\ref{fig:F_px1_py0_pz0}
the pion form factor for two dynamical pion masses.  Fits of the data
to the monopole form of Eq.~(\ref{eq:vmd_form}) are also shown, where shaded
regions correspond to jackknife error bands and the central values are given
in the legend.  The data in Fig.~\ref{fig:F_px1_py0_pz0} were computed
with pseudoscalar pion sink operator fixed at momentum $\bvec{p}_f=(1,0,0)$.
We were unable to compute statistically significant form factors
with the same pseudoscalar sink operator for $\bvec{p}_f=(1,1,0)$
by the ratio method, apparently due to the poor overlap of the sink operator
with the higher momentum state.  We expect that an axial vector sink operator
would be a better choice for future calculations.  A comparison
of the ratio method with the fitting method,
also described in \cite{Bonnet:2003pf,Bonnet:2003aa},
should appear shortly \cite{Fleming:2004}.

This work was supported in part by the Natural Sciences and Engineering
Research Council of Canada and by the U.S. Department of Energy under
contract DE-AC05-84ER40150.  Computations were performed on the 128-node
and 256-node Pentium IV clusters at JLab and on other resources at ORNL,
under the auspices of the U.S.\ DoE's SciDAC initiative.

\vspace{-1ex}

\bibliography{main}

\end{document}